\documentclass[12pt,preprint]{aastex}

\slugcomment{To appear in ApJ}

\shorttitle{Mid-Infrared spectra of QSO2s} \shortauthors{Sturm et
al.}

\begin{document}

\title{Mid-infrared Spitzer spectra of
X-ray selected Type\,2 QSOs: QSO2s are not ULIRGs}

\author{
E. Sturm
, G. Hasinger
, I. Lehmann
, V. Mainieri
, R. Genzel
, M.D. Lehnert
, D. Lutz
, and L.J. Tacconi
} \affil{Max-Planck-Institut f\"ur extraterrestrische Physik,
Postfach 1312, D-85741 Garching, Germany} \email{sturm@mpe.mpg.de}

\begin{abstract}
We have performed a spectroscopic study of 7 Type\,2 QSOs using
the mid-infrared spectrometer {\it IRS} on board the {\it Spitzer
Space Telescope}. These are (to our knowledge) the first mid-IR
spectra of X-ray selected QSO2s taken.
The objects have been selected according to their high intrinsic
luminosities and column densities in the X-rays. Their spectra
strongly differ from template spectra of Type\,2 AGN at lower
luminosities. They do not exhibit strong PAH dust emission
features from circum-nuclear star forming regions, typical for
lower luminosity Type\,2 Seyfert galaxies or other previously used
QSO2 templates, such as the (Ultra)luminous Infrared Galaxy
((U)LIRG) NGC\,6240. They also do not show the ice and silicate
absorption features of highly luminous but deeply embedded compact
nuclei seen in some ULIRGs. Instead they reveal a relatively
featureless, rising continuum similar to luminous Type\,1 AGN. We
also find evidence for a 10$\mu$m silicate feature in {\em
emission}. Models of dusty tori in the AGN unification scenario
predict this only for Type\,1 AGN. The ratio of the AGN continuum
luminosity at 6$\mu$m to the absorption corrected 2-10keV X-ray
AGN luminosity is very similar to that found in Seyfert galaxies.
X-ray selected QSO2s are thus characterized by powerful AGN in
hosts with a luminosity due to star formation $\lesssim$
10$^{11}$L$_{\odot}$. The dominance of the AGN light in the mid-IR
spectra of QSO2s together with their flatter spectral energy
distributions (SEDs) places important constraints on models of the
cosmic infrared background and of the star formation history of
the universe.

\end{abstract}

\keywords{galaxies: quasars --- galaxies: active --- galaxies:
evolution
}

\section{Introduction}
\label{s:intro} Intrinsically luminous, but highly obscured active
galactic nuclei (AGN), the Type\,2 QSOs (QSO2s), have long been
sought as a crucial (and necessary) component of AGN unification
theories and of models that explain the cosmic X-ray and infrared
background by the growth of obscured supermassive black holes
throughout cosmic history. In unified models
different types of AGN are postulated to all harbor a central,
accreting massive black hole surrounded by a Broad Line Region
(BLR). The AGN is proposed to be surrounded by dusty obscuring
material, often modeled as a (clumpy) torus, that anisotropically
absorbs or shadows emission from the nuclear region. Further out a
system of gas (and dust) clouds, the Narrow Line Region (NLR),
emits highly excited narrow lines that are ionized by the central
source. Broad-line, Type\,1 AGN in this scheme are AGN where the
line of sight to the nucleus is not blocked. In narrow-line,
Type\,2 systems the line of sight to the nucleus is blocked by the
obscuring torus, and only the narrow line region is directly
visible. Besides orientation, the luminosity/activity level is the
second primary physical parameter differentiating AGN, from Low
Luminosity AGN to Seyferts to QSOs.

Population synthesis models of the cosmic X-ray background (XRB)
try to explain the hard spectrum of the XRB by a mixture of
absorbed and unabsorbed AGN, spread over a large redshift interval
(e.g. Setti \& Woltjer 1989), folded with the corresponding
luminosity function and its cosmological evolution. Such models
were developed within the unified AGN scenario, often assuming a
fraction of obscured (Type\,2) to unobscured (Type\,1) objects as
in local AGN (ascribed to the covering factor of the torus which
is assumed to be the same for all
luminosites). They require that a significant fraction of the
cosmic black hole growth is obscured by large amounts of dust and
gas. Hence, a substantial contribution to the XRB should come from
intrinsically luminous, obscured X-ray sources, i.e. QSO2s.

In the Seyfert domain unification theories are very successful.
The existence of obscured QSOs, however, as the high-luminosity
analogues of the Seyfert 2 galaxies, has been hotly debated since
the formulation of the first unified theories. Radio-loud QSO2s,
i.e. radio galaxies, have been the only observed Type\,2 QSOs for
a long time. However, radio galaxies represent only a small
fraction of the total AGN population. Radio quiet QSO2s remained
elusive until very sensitive and deep hard x-ray imaging and
spectroscopy campaigns became feasible with the advent of Chandra
and XMM-Newton (Dawson et al. 2001, Norman et al. 2002, Mainieri
et al. 2002, Stern et al. 2002, Hasinger et al. 2003, Della Ceca
et al. 2003). In the X-ray domain the QSO-2 population is
characterized by high intrinsic absorption (N$_H$ $>$ 10$^{22}$
cm$^{-2}$), and high intrinsic X-ray luminosity (L[0.5-10keV] $>$
10$^{44}$ erg/s).

According to both AGN unification and XRB models, QSO2s should be
more numerous than Type\,1 QSOs. From XMM-Newton observations of
the Lockman Hole we (Mainieri et al. 2002) found indeed a large
fraction of the objects to be obscured AGN. However, the number of
identified {\em luminous} sources, i.e. QSO2s, amongst them is
very low compared to the model assumptions. The fraction of
Type\,2 objects in these surveys decreases with increasing X-ray
luminosity (Ueda et al. 2003, Hasinger 2003). A similar trend has
been found in optical studies based on QSOs from the SDSS (Simpson
2005). This trend might indicate a breakdown of those unification
models where the covering factor is independent of luminosity. An
alternative explanation might be that many QSO2s are so obscured
that they are hidden even at hard X-rays (`Compton thick', N$_H$
$>$ 10$^{24}$ cm$^{-2}$). Recently, deep mid-infrared imaging
studies with Spitzer, using mid-IR/radio color criteria, have
begun to detect a population of Type\,2 QSOs which seem to be at
least comparable in number density to the un-obscured Type\,1
population (Martinez-Sansigre et al. 2005, Donley et al. 2005).
Some of these objects show no hard X-ray counterpart, indicating
that these QSOs are hidden by Compton thick material even at high
X-ray energies.

In this paper we present the first mid-infrared spectra of QSO2s.
The objects have been selected according to their X-ray luminosity
and column density (section \ref{s:sample}). The spectra and the
multi-wavelength spectral energy distributions (SEDs) of our QSO2s
allow us to discuss the mid-infrared properties of such QSO2s by
comparing them to various template spectra of active galaxies
(section \ref{s:results}). We study the relative contributions
from star formation and obscured AGN to their total energy output
(section \ref{s:powersource}), and we investigate the putative
luminosity dependence of the relative abundance of Type\,2 vs.
Type\,1 objects (section \ref{s:xtoir}).

\section{Sample Selection, Observations, Data Processing}
\label{s:sample} We have selected the QSO2s for this program from
our deep ($\approx$ 1Msec) XMM-Newton observations of the Lockman
Hole (Hasinger et al. 2001, Mainieri et al. 2002). We picked those
objects with high intrinsic absorption (with a mean N$_H$
$\approx$ 1.0x10$^{23}$ cm$^{-2}$, as derived from absorbed power
law fits), and high intrinsic (absorption corrected) X-ray
luminosities (L[0.5-10keV] $>$ 10$^{44}$ erg/s). From this list of
objects we culled the ones which, in addition, have ISOCAM
detections with a flux density at 15 $\mu$m of greater than 0.3
mJy (Fadda et al. 2002). As a control target in order to examine
the role of luminosity in this selection method we have also
selected one source (LH901A) with similarly high X-ray absorption,
but slightly lower intrinsic luminosity. Furthermore, we have
performed optical spectroscopy on our targets, which yielded
narrow emission lines typical of Type\,2 AGN without associated
broad components (Lehmann et al. 2000, 2001).

We have augmented the sample by 3 targets with similar information
from the literature: the QSO2 `proto-type' CDF-S 202 (CXOCDFS
J033229.9-275106) from the Chandra Deep Field-South (Norman et al.
2002), CXO 52 (CXO J084837.9+445352, Stern et al. 2002), and AX
J0843+2942 (della Ceca et al. 2003). The latter source is a radio
loud QSO2 candidate, which we chose to investigate the radio loud
end of the QSO2 phenomenon. These 8 sources also cover a large
redshift range from 0.2 to 3.7. The targets are listed in Table
\ref{tab:targets}.

Our data were obtained with the Spitzer IRS (Houck et al. 2004;
Werner et al. 2004) as a GO Cycle-1 project (PI: E. Sturm, PID
3223). Because of the relative faintness of our objects they were
observed in low resolution mode only, restricted to the one or two
IRS slits and orders, which (together) covered a rest wavelength
range of approximately 5 to 12$\mu$m. Exposure times were
typically between 4800 and 8400 seconds in the LL module (20 to 35
nod1-nod2 cycles of 120s at each nod position), and between 3840
and 7200 seconds in the SL module (8 to 15 cycles of 240s per
nod).

Our data reduction started with the two-dimensional BCD products
from the Spitzer pipeline (S12). We removed the sky background by
a pairwise subtraction of frames at different nod positions for
each cycle. Using the pipeline bad pixel mask (bmask) and sigma
clipping algorithms we removed (de-glitch) bad or noisy pixels,
replacing them by an interpolation of the neighboring pixels. The
resulting nod1-nod2 frames were then co-added by averaging,
resulting in a single, filtered, frame with a positive and a
negative beam. We used the software package SMART (Higdon et al.
2004) to extract calibrated one dimensional spectra for the
positive and negative beams (using the `interactive/column/No
SkySub' method). These were then averaged into the final spectra.

In the case of LH\,H57 a second object appears in the LL2 slit,
which partially overlaps with LH\,H57.  This second object is
located approximately 10\arcsec (2 pixels) ~SE of LH\,H57
(PA$\approx$125$^\circ$ E of N). It is therefore not visible in
the SL1 slit, which has a smaller slit width and a different
orientation. We have extracted the LL2 spectra of both objects by
using the `fixed column' method in SMART. This has caused some
flux loss (and increased flux calibration uncertainty) for either
of the two spectra, but allowed a qualitative separation of the
two objects. The second object appears to be starburst dominated,
and we tentatively assign a redshift of z $\approx$ 0.58 for that
second source, by identifying the two major peaks in the spectrum
with the 11.3 and 12.7 $\mu$m PAH features.

\section{Mid-infrared spectral properties and global SEDs}
\label{s:results} 
\begin{figure}[t]\epsscale{1.0} \plotone{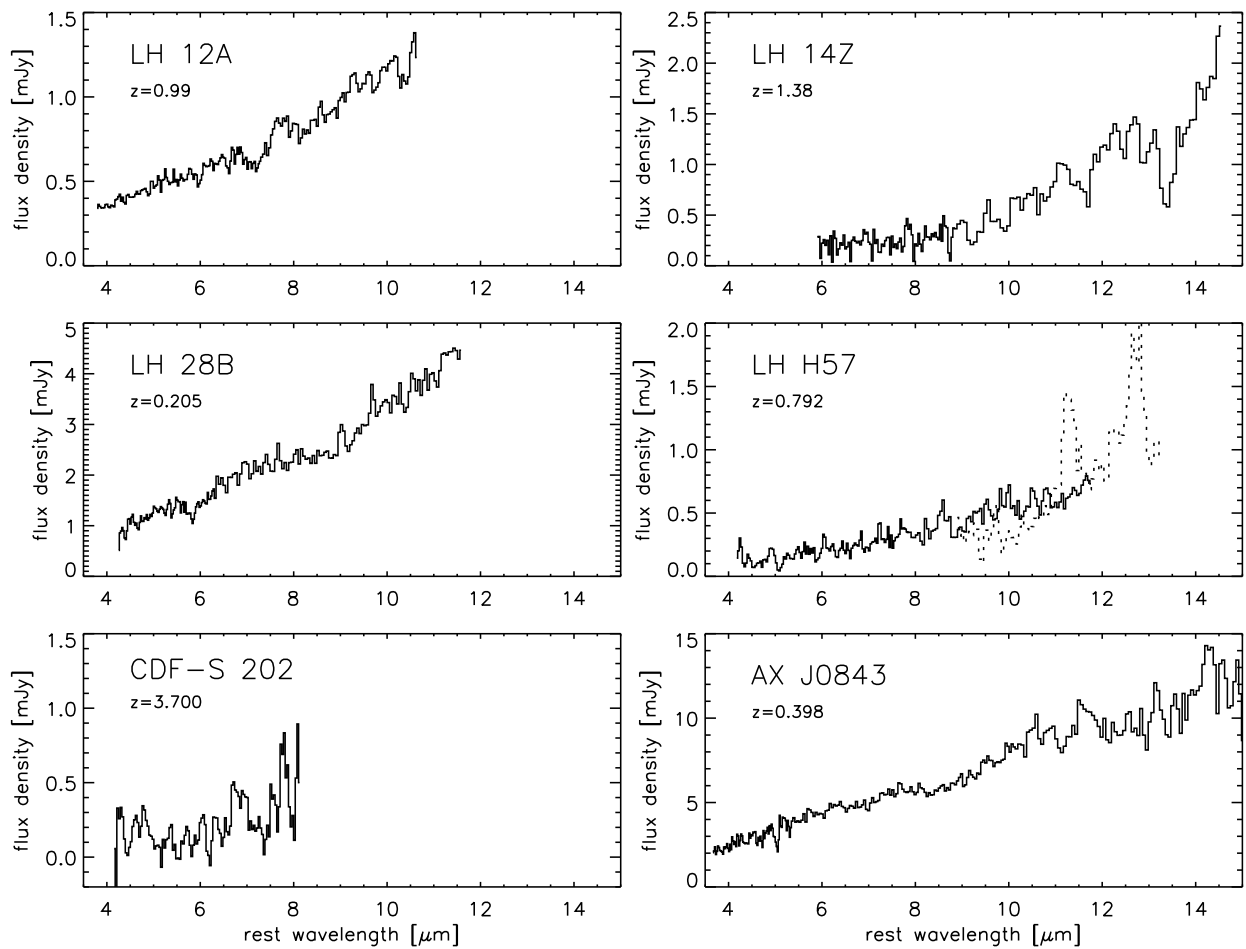}
\caption{IRS spectra of 6 of the 7 QSO2s. CXO 52 Lynx was not
detected.  The dotted line in the LH\,H57 panel is the spectrum of
the second source in the LL2 slit which is partly blended with
LH\,H57, assuming z=0.579. } \label{F:spectra}
\end{figure}

\begin{figure}[t]\epsscale{1.0} \plotone{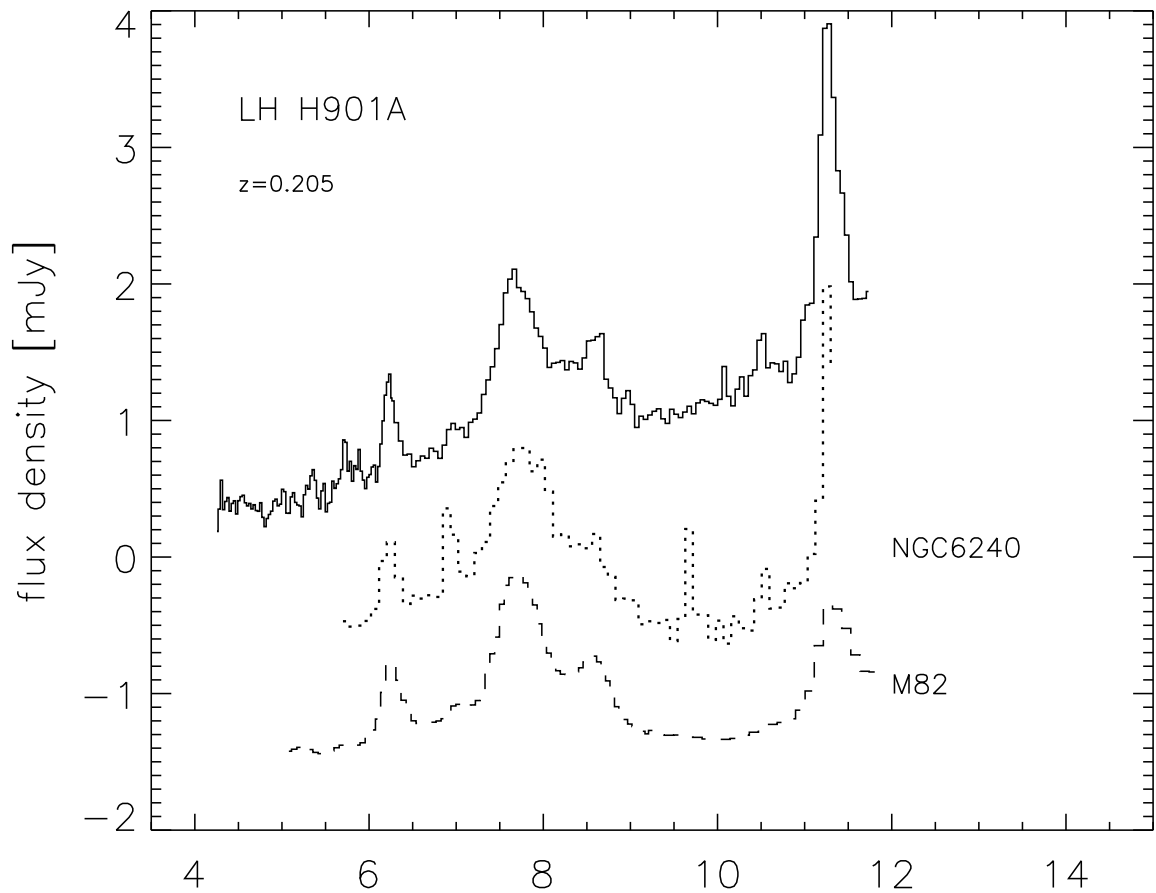}
\caption{IRS spectrum of LH 901A. The dotted and dashed lines are
the ISOPHOT-S spectrum of NGC\,6240 (Lutz et al. 2003) and the
ISOCAM-CVF spectrum of M\,82 (F\"orster Schreiber et al. 2003),
arbitrarily scaled.} \label{F:LH901a}
\end{figure}

\begin{figure}[t]\epsscale{1.0} \plotone{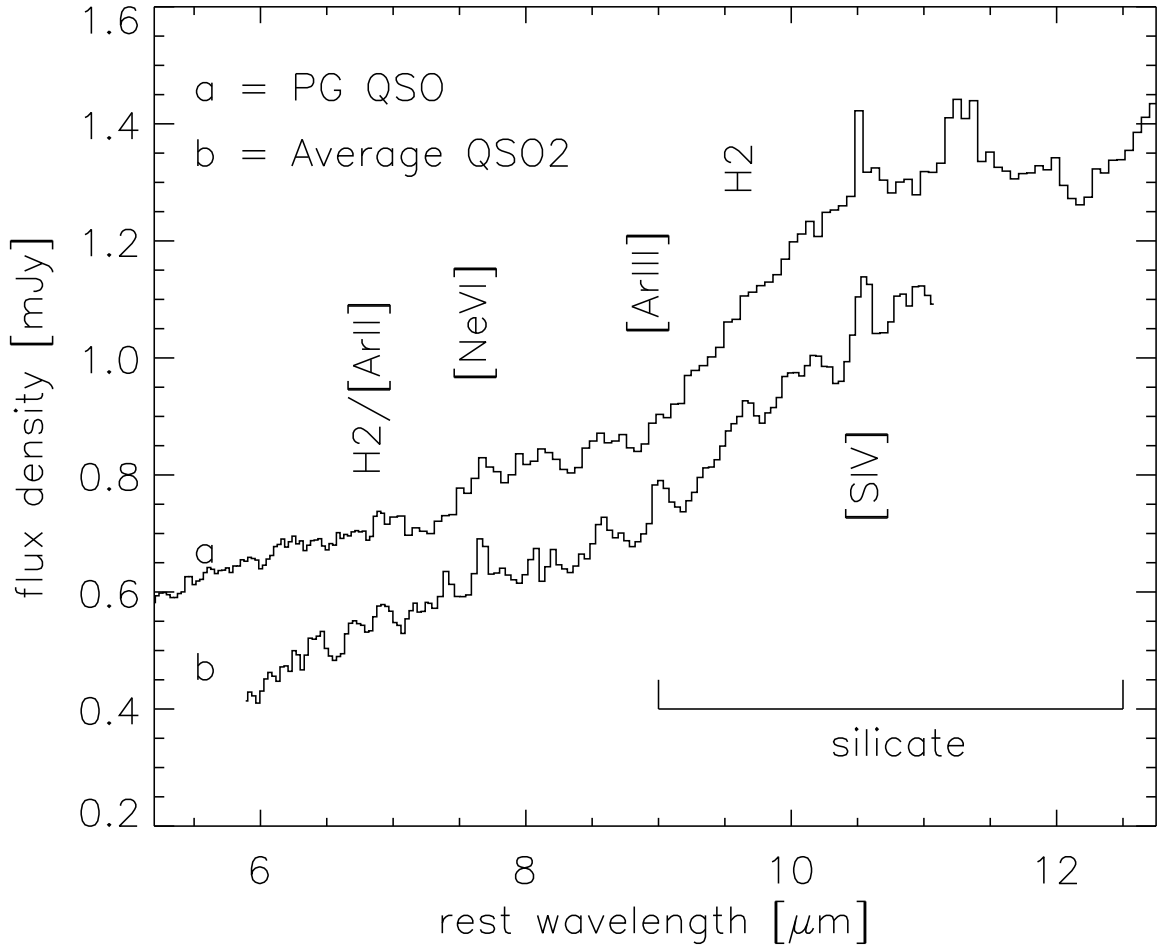}
\caption{{\it a (top):} a typical PG QSO (PG1426+015, arbitrarily
scaled), {\it b (bottom):} the average QSO2 spectrum.}
\label{F:average}
\end{figure}

The spectra are shown in Fig. \ref{F:spectra}
and Fig. \ref{F:LH901a}. With the exception of CXO 52 Lynx all
objects were detected. We derive an upper limit for CXO 52 in the
5 - 9 $\mu$m rest (20 - 40 $\mu$m observed) wavelength range of
0.1 mJy. All the spectra show a rising mid-infrared continuum,
typical of AGN. There are no strong hints of absorption from
silicates (9.6 $\mu$m), ices (6 $\mu$m) or hydrocarbons (6.8
$\mu$m), which would be typical of highly obscured objects. Dust
emission features (PAHs) from star forming regions are very weak
or absent, with the obvious exception of LH\,901A (Fig.
\ref{F:LH901a}), the object with the lowest intrinsic X-ray
luminosity of our sample. No strong ionic or molecular (narrow)
emission lines can be seen in these faint, low resolution spectra.
Only LH\,901A and LH\,28B show weak lines of [S\,IV], [Ar\,III]
and H$_2$. We have computed an average spectrum (scaled with the
6$\mu$m flux) using LH 12A, 14Z, 28B, H57, and AXJ0843, which is
shown in Fig. \ref{F:average}. In this average spectrum we
identify (unresolved) narrow emission lines of [Ar\,II]6.9$\mu$m
(probably blended with H$_2$\,S(5)), [Ar\,III]9.0$\mu$m
(potentially blended with [Mg\,VII]), [Ne\,VI]7.6$\mu$m, and
H$_2$\,S(3)9.6$\mu$m. For comparison we show in Fig.
\ref{F:average} the IRS spectrum of a luminous Type\,1 AGN, the
QSO PG1426+015, a typical PG QSO from a large study (QUEST) of
ULIRGs and PG QSOs (Veilleux et al., in preparation). This
comparison supports the conclusion that the spectra of our objects
exhibit continuous spectra like luminous Type\,1 AGN in the
mid-infrared regime, despite being selected as highly obscured,
Type\,2 AGN in the X-rays.

This absence of PAH and/or absorption features is remarkable and
surprising. QSO2s are often believed to be related to
Ultraluminous Infrared Galaxies (ULIRGs). ULIRGs are extremely
dusty, highly obscured galaxies with space densities and
luminosities comparable to local QSOs (Sanders \& Mirabel 1996).
In particular NGC\,6240, a nearby infrared luminous system, is
frequently used as type 2 AGN template (e.g. Lehmann et al. 2002,
Norman et al. 2002) because its X-ray and mid-infrared properties
identify it as hosting a highly obscured AGN (Lutz et al. 2003,
and references therein). The mid-IR spectrum of NGC\,6240,
however, exhibits strong dust emission features quite different
from our QSO2 spectra (but very similar to LH901A, see the
comparison in Fig. \ref{F:LH901a}). Some local ULIRGs, like IRAS
F00183-7111 (Tran et al. 2001, Spoon et al. 2004), and many
mid-infrared sources at higher redshifts (Houck et al. 2005, Yan
et al. 2005), are dominated by absorbed continua, shaped by deep
absorption features of silicates, ices, and hydrocarbons,
indicating compact, deeply embedded nuclei that could plausibly be
AGN. These sources are alternative template candidates for QSO2s.
However, the objects presented here clearly do not resemble these
dust absorbed ULIRGs.

Even more remarkable, the average QSO2 spectrum appears to exhibit
a 10$\mu$m silicate feature in {\em emission}, even if our spectra
do not extend far enough beyond 10$\mu$m to cover the full
emission feature. The PG QSO shown in Fig. \ref{F:average} covers
a broader range and exhibits a clear 10$\mu$m silicate emission
feature. From the close similarity of the two spectra we conclude
that the average QSO2 spectrum shows a comparable emission
feature. Silicate emission was recently found in many PG QSOs
(Siebenmorgen et al. 2005, Hao et al. 2005, Sturm et al. 2005).
Many AGN torus models predict such a silicate emission as a
signature of the obscuring torus. According to unification models
silicates should appear in emission in Type\,1 AGN, but in
absorption in Type\,2 AGN. Sturm et al. (2005) have urged caution
that this may not be the full explanation for the observed
silicate emission in QSOs, and that at least some emission may
arise in more extended regions, like a dusty narrow line region.
They substantiate this claim with the relatively cool (~200K)
temperature of the silicate dust (as derived from the ratio of the
emission features at 10 and 18 $\mu$m), which seems too cold to be
explained by a hot inner torus wall, but which is consistent with
observed color temperatures of extended dust found in the narrow
line regions of some nearby Seyfert galaxies. The presence of
silicate emission in Type\,2 QSOs, if confirmed in future
observations, would pose further important constraints on the
origin of dust emission in AGN.

\begin{figure}[t]\epsscale{0.9} \plotone{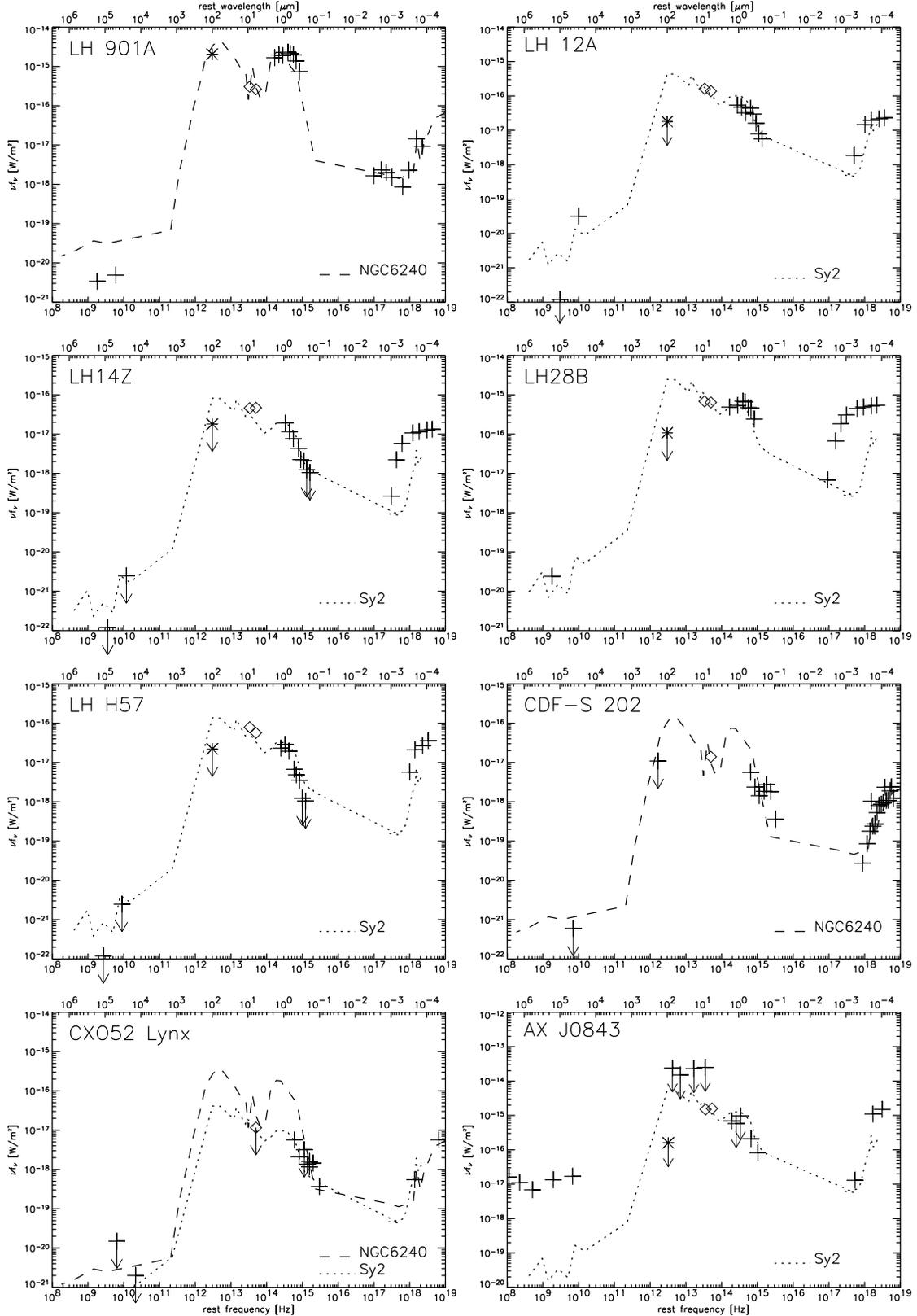}
\caption{SEDs of the QSO2s. Diamonds represent our IRS spectra at
6 and 9 $\mu$m rest wavelength. {\it Dashed}: NGC\,6240, {\it
dotted}: average Sy2, scaled to the X-ray or near-IR. Asterisks at
the SED peak are predictions from the PAH strength (see text).}
\label{F:SEDs}
\end{figure}

We have collected the multi-wavelength (X-ray-to-radio) SEDs of
our objects, using our ancillary data and data from the literature
(Norman et al. 2002, Stern et al. 2002, della Ceca et al. 2003).
In Figure \ref{F:SEDs} we compare these SEDs to the SEDs of
NGC\,6240 and an average Seyfert\,2 SED, composed from a sample of
local galaxies by Schmitt et al. (1997, soft X-ray to radio range)
and Moran et al. (2001, hard X-ray range). The average Seyfert 2
spectrum has a lower infrared/X-ray ratio than NGC\,6240 (see
Figure \ref{F:SEDs}, lower left panel). For most objects we have
scaled the comparison SEDs to the X-ray data. For LH14Z, 28B, and
H57, which are unusually X-ray bright, we have scaled to ´the
near-IR data. In order to predict a flux density at 100$\mu$m, the
wavelength around which the SEDs of many galaxies have their peak,
we have measured 3$\sigma$ upper limits for the PAH 7.7$\mu$m peak
flux densities (above the continuum). This PAH peak height
(S$_{7.7}$) is a reasonable tracer of the star formation related
far-IR flux of active galaxies (e.g. Lutz et al. 2003). In M\,82,
a prototypical starburst galaxy (shown in Fig. \ref{F:LH901a}),
the ratio of the 7.7$\mu$m PAH peak height to the 100$\mu$m flux
density (S100), both measured in Jy, is 0.069 (derived from the
ISOCAM-CVF spectrum, F\"orster Schreiber et al. 2003, and the
ISO-LWS spectrum, Colbert et al. 1999). This ratio is very similar
in many starburst galaxies, with a dispersion of about 35\%
(Rigopoulou et al. 1999). Using this value we predict 100$\mu$m
flux density limits for the QSO2s. They are listed in Table
\ref{tab:targets} (denoted as S100$_{PAH}$) and indicated as
asterisks in Figure \ref{F:SEDs}. The optical-to-near-IR SEDs of
most of our objects are quite similar to the average Seyfert\,2
galaxy, or to NGC\,6240. At these wavelengths (and within the
redshift range of our objects) the emission of Type\,2 quasars is
probably dominated by starlight from the host galaxy. The
mid-to-far infrared SEDs of these QSOs, however, are clearly
flatter (have smaller mid-far-IR/near-IR ratios) than normal
Type\,2 template objects, if our PAH based extrapolation is
applicable. Only LH901A, with its intrinsic X-ray luminosity more
in the range of normal Seyferts than of QSOs, matches quite nicely
the NGC\,6240 template. The mid-IR SEDs of QSO2s are therefore on
average warmer than for Type\,2 template objects of lower
luminosity, probably with only a weak (rest-frame) far-IR peak,
due to a strong AGN contribution as revealed by the mid-IR
spectra. We estimate the star formation related far-IR
(8-1000$\mu$m) flux, F$_{IR,SF}$, and luminosity, L$_{IR,SF}$,
again from the measured PAH peak flux density limits, applying a
ratio of log(S$_{7.7}$/F$_{IR,SF}$) = 12.0, as measured in M\,82.
The corresponding values of L$_{IR,SF}$ for our targets are listed
in Table \ref{tab:targets}. With values (limits) of a few x
10$^{10}$L$_\odot$ they lie in a range which is very typical for
local starburst galaxies but well below ULIRGs.

Our decision to select only those objects with a (estimated)
15$\mu$m flux density above 0.3mJy might have caused some
selection effect, favoring AGN-like objects with strong mid-IR
continua. We note, however, that the 0.3 mJy level does not impose
a severe cut-off, in the sense that at such a flux density we
would have detected PAH dominated objects like NGC\,6240, too.
NGC\,6240 itself would be detectable with {\it IRS} out to a
redshift of $\sim$ 1.5. If CXO\,52 were NGC\,6240-like (but at
higher absolute flux level, scaling with the X-ray luminosity) we
would have detected it (see Fig. \ref{F:SEDs}).

\section{The relative contributions of star formation and AGN}
\label{s:powersource} A quantitative estimate of the contribution
of obscured AGN to the X-ray and infrared backgrounds is crucial
to deduce the star formation history of the universe from galaxy
luminosity functions (see section \ref{s:intro}), and it is
important to understand the evolution of galaxies and AGN. In
contrast to their lower luminosity AGN cousins, the Seyfert\,2
galaxies, contributions of the host galaxies to the mid-IR spectra
of QSO2s seem to be small, since typical star-formation tracers
such as PAH emission features are mostly absent. In order to
quantify this we have performed a decomposition of the spectra
into a starburst template (M\,82) and an AGN template (linear
continuum) in the same way as described in Lutz et al. (2004): in
the range covered by our IRS spectra the AGN emission is best
isolated shortward of the complex of aromatic emission features
(Laurent et al. 2000). We determine a continuum at 6$\mu$m rest
wavelength, and eliminate non-AGN emission. This is done by
subtracting the PAH template scaled with the strength of the
aromatic features arising in the host or in circum-nuclear star
formation.

The resulting 6$\mu$m AGN continua are listed in Table
\ref{tab:targets}. The AGN contribution to the total flux density
at 6$\mu$m is 67\% for the lower luminosity source LH901A, and
close to 100\% for the QSO2s (except CDF-S202 and CXO 52 where we
could not measure it). QSO2s likely represent a significant
component of the high redshift AGN population. Since their
infrared spectra are more AGN dominated (i.e. flatter) than less
luminous Type\,2 AGN, their contribution to the peak of the cosmic
infrared background is lower than predicted using these less
luminous templates. Future models of the star formation history
and the cosmic infrared background have to take this into account.
The flatter infrared SEDs of these QSO2s also explain (at least
partially) the low success rate of previous attempts to detect
these objects in the (sub-)mm with SCUBA and MAMBO (e.g. Chapman
et al. 2004), if we assume a similar 100$\mu$m/850$\mu$m flux
density ratio as in the lower luminosity templates.

\section{The relation between mid-infrared
continuum and hard X-ray emission}
\label{s:xtoir}

Hard X-rays, unless extremely absorbed in Compton-thick objects,
can provide a direct view to the central engine. They are,
therefore, a measure of the bolometric luminosity of the AGN. The
nuclear infrared continuum in AGN, in contrast, is due to a
re-processing of the AGN emission by the circum-nuclear dust, e.g.
in the putative torus. The observed mid-infrared AGN emission is
thus a function of both the AGN luminosity and the distribution of
the obscuring matter. As mentioned in the introduction one
important result of the hard X-ray studies is the decreasing
fraction of Type\,2 objects with increasing X-ray luminosity. A
possible explanation of this trend may be that high luminosity
objects are able to 'clean out' their environment by ionizing the
circumnuclear matter and/or producing strong outflows, while low
luminosity objects are largely surrounded by the circumnuclear
starburst region from which they are being fed. A similar scenario
is provided by the `receding torus' models (e.g. Lawrence 1991,
Simpson 2005). In these models the opening angle of the torus
(measured from the torus axis to the equatorial plane) is larger
in more luminous objects because the distance of the inner torus
wall to the nucleus, which is determined by the dust sublimation
temperature, increases with luminosity (while the torus height
stays constant). If this trend of decreasing Type\,2 AGN fraction
with increasing luminosity is indeed related to a decreasing
covering factor the ratio of mid-IR continuum (re-radiation) and
intrinsic X-ray emission should be lower in QSO2s than in
Seyfert2s.

Lutz et al. (2004) have examined this ratio of hard X-ray to
mid-IR continuum luminosities in a large sample of local Seyfert
galaxies (and a few QSO1s), based on the decomposition of their
mid-IR spectra into AGN and starburst components as described
above. They found a good correlation (with some scatter) and no
difference between Type\,1 and Type\,2 objects. In Table
\ref{tab:targets} we list the X-ray/mid-IR ratios of our QSO2s,
derived in the same way as in Lutz et al. (2004). The ratios are
very similar (within the dispersion) to the ratios in Seyfert
galaxies, while a decreasing covering factor would cause an
increase of the relation at high luminosities. However, given the
scatter in our data and the variability of the X-ray data small
changes in covering factor would be difficult to detect. Our data
are thus consistent with a constant Type\,2/Type\,1 ratio, but we
cannot exclude modest luminosity dependencies of the covering
factor at this stage. On the other hand, the observed X-ray/mid-IR
correlation offers an explanation for the overall mid-IR spectral
properties of QSO2s: the AGN continuum at mid-IR wavelengths
simply scales with the high AGN luminosities in our objects and
outshines any contribution from circum-nuclear star forming
regions ($\lesssim$ 10$^{11}$L$_\odot$, see section
\ref{s:results}). The reason for the constancy of this relation
and the similarity of Type\,1 and Type\,2 AGN in that respect
remains an open question. Lutz et al. (2004) have tentatively
attributed this similarity to extended dust emission, because
significant non-torus contributions to the AGN mid-IR continuum
could mask the expected difference between the two types of AGN.
Our QSO2 study, which extends this similarity of Type\,1 and
Type\,2 AGN to the quasar regime, and which indicates silicate
emission (rather than absorption) in Type\,2 QSOs, has added
evidence that at least some of the mid-IR dust emission in AGN
arises in extended regions.

\acknowledgments

This work is based on observations made with the Spitzer Space
Telescope, which is operated by the Jet Propulsion Laboratory,
California Institute of Technology under NASA contract 1407. We
are grateful to Hagai Netzer for inspiring discussions.

\clearpage

\begin{table*}[t]
\scriptsize \caption{The QSO2 sample, X-ray, and infrared
properties. \label{tab:targets}}
\begin{tabular}{lcccllllll}
\tableline\tableline \bf \rule[-0.1in]{0in}{0.3in} Source &  RA &
DEC & z & log L$_x$ & log N$_H$ & S$_{AGN,6}$ &
L$_x$/$\nu$L$_{\nu,6}$ & S100$_{PAH}$ & L$_{FIR,SF}$\\
            & \multicolumn{2}{c}{Equatorial J2000}     & & & & [mJy]& & [mJy] & [10$^{10}$L$_\odot$]\\\hline
LH 12A      & 10h51m48.80s & +57d32m48.0s & 0.990      & 44.3      & 22.7      & 0.58 & 0.21 & $<$0.6  & $<$2.5\\
LH 14Z      & 10h52m42.20s & +57d31m58.0s & 1.38       & 44.9      & 22.6      & 0.20 & 0.36 & $<$0.6  & $<$4.9 \\
LH 28B      & 10h52m50.30s & +57d25m44.0s & 0.205      & 44.2      & 21.3      & 1.63 & 1.39 & $<$3.4 & $<$0.5\\
LH 901A     & 10h52m52.80s & +57d29m00.0s & 0.205      & 43.1      & 23.5      & 0.45 & 0.32 &  14.4   & 2.2\\
LH H57      & 10h53m05.70s & +57d28m10.3s & 0.792      & 44.0      & 22.4      & 0.20 & 0.32 & $<$0.8  & $<$1.9 \\
CDF-S 202   & 03h32m29.86s & -27d51m05.8s & 3.700      & 45.0$^{1)}$  &$>$24   & 0.18 & 0.23 &  - & - \\
CXO 52 Lynx & 08h48m37.90s & +44d53m52.0s & 3.288      & 44.5$^{1)}$  & 23.7   & $<$0.1 & $<$0.2 &  - & -\\
AX J0843+2942 & 08h43m09.90s & +29d44m04.9s & 0.398    & 45.5      & 23.2      & 4.20 & 2.67 & $<$5.1 & $<$3.0\\
\hline \tableline
\end{tabular}
\newline
col(1): target name; col(2): RA; col(3):DEC; col(4):redshift;
col(5): intrinsic luminosity L[0.5-10keV] in erg/s; col(6):
intrinsic absorption log(N$_H$) in cm$^{-2}$, col(7): AGN
continuum at 6$\mu$m in [mJy], estimated from spectral
decomposition; col(8): intrinsic 2-10keV luminosity vs. AGN
continuum luminosity ($\nu$L$_\nu$(6$\mu$m); col(9): predicted
(upper limit of the) flux density at 100$\mu$m in [mJy], derived
from 7.7$\mu$m PAH strength/limit; col(10): star formation related
8-1000$\mu$m luminosity, estimated from 7.7$\mu$m PAH \\
1) 2-10 keV\\
\end{table*}

\end{document}